\documentclass[runningheads,a4paper]{llncs}

\usepackage{amssymb}
\usepackage{rotating}
\usepackage{graphicx}
\usepackage{listings}
\usepackage{tabulary}
\usepackage{tabularx,booktabs}

\usepackage{url}
\urldef{\mailsa}\path|{rrychtarikova, stys}@frov.jcu.cz|
\newcommand{\keywords}[1]{\par\addvspace\baselineskip
\noindent\keywordname\enspace\ignorespaces#1}

\begin{document}

\mainmatter
\title{Observation of dynamics inside an unlabeled live cell using bright-field photon microscopy: Evaluation of organelles' trajectories}
\titlerunning{Dynamics inside an unlabeled live cell}
\author{Renata Rycht\'{a}rikov\'{a}\and Dalibor \v{S}tys}

\authorrunning{Rycht\'{a}rikov\'{a} \& \v{S}tys}
\institute{University of South Bohemia in \v{C}esk\'{e} Bud\v{e}jovice,\\ Faculty of Fisheries and Protection of Waters,\\ South Bohemian Research Center of Aquaculture and Biodiversity of Hydrocenoses,\\ Institute of Complex Systems,\\ Z\'{a}mek 136, 373 33 Nov\'{e} Hrady, Czech Republic\\
\mailsa\\
\url{http://www.frov.jcu.cz/cs/ustav-komplexnich-systemu-uks}}

\toctitle{Lecture Notes in Bioinformatics}
\tocauthor{Rycht\'{a}rikov\'{a} \& \v{S}tys}
\maketitle

\begin{abstract}
This article presents an algorithm for the evaluation of organelles' movements inside of an unmodified live cell. We used a time-lapse image series obtained using wide-field bright-field photon transmission microscopy as an algorithm input. The benefit of the algorithm is the application of the R\'{e}nyi information entropy, namely a variable called a point information gain, which enables to highlight the borders of the intracellular organelles and to localize the organelles' centers of mass with the precision of one pixel.   
\end{abstract}

\section{Introduction}

The observation of cell biophysics such as organelles' movements plays a pivotal role in the evaluation of physiological state of mammalian cells, e.g. in recognition of the cell pathology. This observation can be performed using bright-field transmission is a classical light-microscopic method, whose advantage is a possibility of observation of an unlabeled live cell and tissues. It also does not require complicated sample preparation.

However, the disadvantage of this microscopic technique is a low contrast of the structures observed in transparent samples due to the interferences of the light which passes the optical path of the microscope. In order to highlight the intracellular structures' borders in digital images, we proposed a novel information-entropic variable derived from the R\'{e}nyi information entropy -- a point information gain. In the image processing, beside the others benefits \cite{rychtarikova2016a,rychtarikova2016b,rychtarikova2016c}, this variable enables to work with intensity histograms and to calculate one resulted value of the point information gain for similar intensity values.

In the presented algorithm, this approach is employed to localize centers of mass of dynamical light-diffracting objects in order to track them in time-lapse images of a mammalian cancer cell. These trajectories are further normalized to the scanning time and statistically analyzed.

\section{Results and Discussion}

\subsection{Description of Image Processing Algorithm} \label{algorithm}

The whole time-lapse raw file series scanned using the nanoscope~(Sect.~\ref{Microscopy}) was processed as follows:

\begin{enumerate}
\item Using Matlab$^{\circledR}$ software fortified by an Image Processing Toolbox and Statistics Toolbox (Mathworks, USA), the image of the cell was got rid of the background using a cumulative mask (Fig.~\ref{Fig1}), where the contribution to the mask was created from the absolute values of the difference between the green intensities in the original series image and in the blurred version of this image (using a 10-px disk-shaped image filter). The green intensities in the original image were calculated as arithmetic average of each two green pixels in Bayer filter quadruplets~\cite{bayer} which gave a quater-sized image compared with the original raw data~\cite{tkacik}.

\begin{figure}
 \centering 
\includegraphics[width=0.75\textwidth]{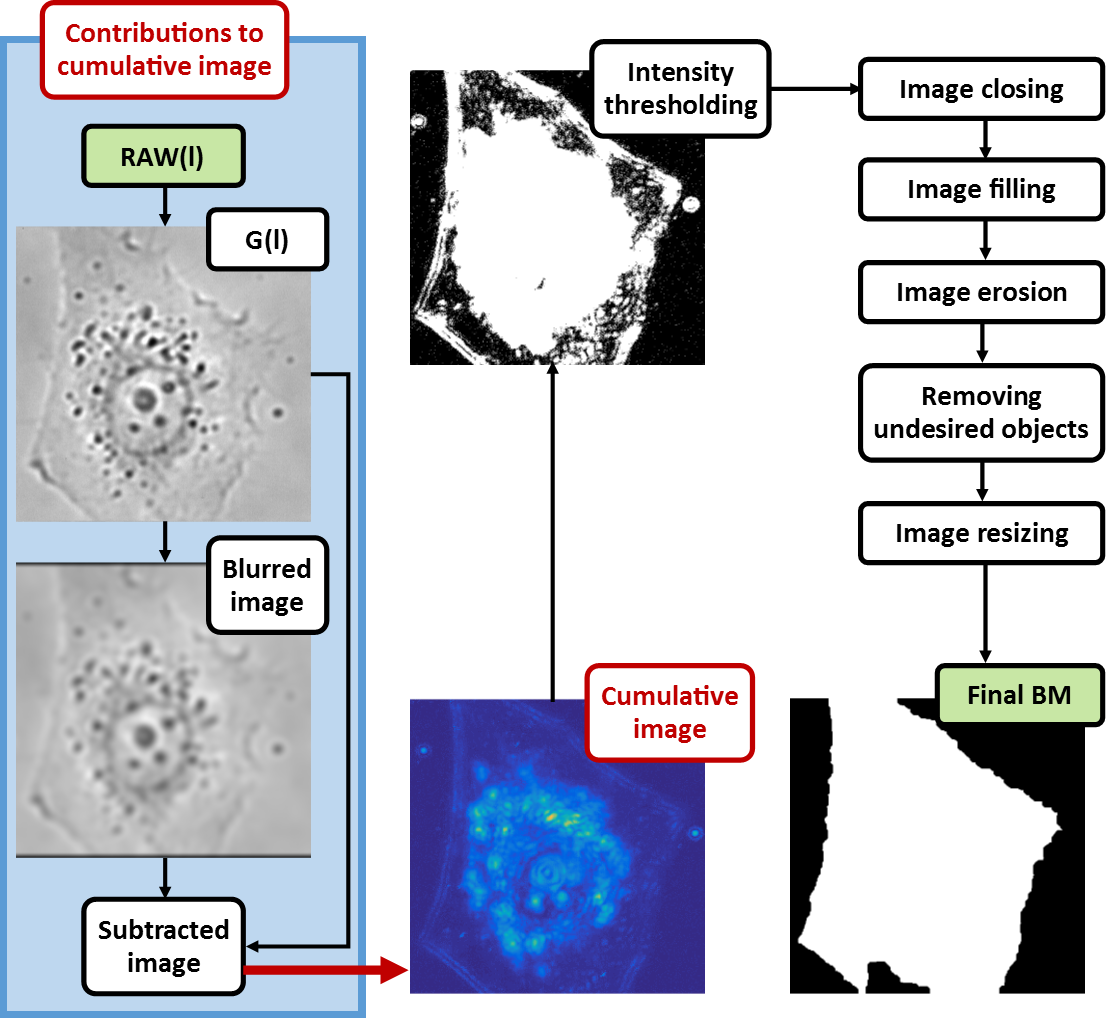}
 \caption{Algorithm for segmentation of a cell from time-lapse series obtained using wide-field bright-field photon transmission microscopy.} 
\label{Fig1}
\end{figure}

The obtained mask was further binarized and processed by a standard image-processing methods -- by thresholding of the relevant intensities, morphological closing (a 3-px disk-shaped structuring element), filling the holes in the binary image, and morphological erosion to remove the blurred edges of the cell (a 10-px disk-shaped structuring element). The resulted binary mask was two-fold increased to the size of the original raw file and applied to the whole time-lapse series to select the cell of interest at each time step. 

Positions of green pixels darker than the intensity mode of the background surrounding the cell in each series image was stored in a .mat file and used for next image processing (see Item 3).  

\item For each segmented raw file in the series, the values of point information gain ($\Gamma_{\alpha}^{(i)}$, [bit]) were computed using the Image Info Extractor Professional software (Institute of Complex Systems FFPW USB) at omitting black (zero) pixels surrounding the cell of interest as

\begin{equation}\label{Eq1}
\Gamma_{\alpha}^{(i)} = \frac{1}{1-\alpha}\log_{2}\frac{\sum_{k=1}^n{(p_k^{(i,j)})^\alpha}}{\sum_{k=1}^n{p_k^\alpha}},
\end{equation}
where $\alpha$ is a dimensionless R\'{e}nyi coefficient. For red and blue pixels of the raw data, the probabilities $p_j$ and $p_j^{(i)}$ characterize relative frequencies of occurrences of $k$ occupied intensities in the histogram of relevant color pixels and in the same histogram where one element of the bin $i$ is missing, respectively. In case of the green pixels, two elements at the positions $i,j$ were removed from the histogram. For each color image channel, the output data were stored as an 8-bit color image where black and white pixels corresponds to the $\Gamma_\alpha^{(i)}$ with highest and lowest frequencies of occurrences in the raw data.

\item Algorithm in Fig.~\ref{Fig2} implemented in Matlab$^{\circledR}$ software (see Sect.~\ref{supplementary}) was used for tracking light-diffracting organelles inside the images of cell. Through the whole the time-lapse series, the zero values (black pixels) in the green channel of each 8-bit $\Gamma_{4.0}^{(i)}$- (i.e., PIG-) image were thresholded. The number of white pixels in the cell interior were further reduced by selection of the intensities darker than original cell's background (see Item 1). In order to segment only relevant large organelles in the binary image, white structures connected to image border and demarcated the cell interior as well as small light structures were suppressed. So-selected organelles (Fig.~\ref{Fig3}--\textit{right}) as connected components in binary image were labeled to give a number matrix~$L$.     

The main body of the algorithm for tracking the organelles started by labeling of the objects (i.e., light-diffracting organelles) and calculation of their total number in the 1$^{\mbox{st}}$ PIG-image of the series and continued by two loops. In the higher-order for-loop, each organelle of the 1$^{\mbox{st}}$ label matrix $L_1$ and the label matrix itself were renamed. Starting the 2$^{\mbox{nd}}$ PIG-image of the time-lapse series, the organelle was traced through the above-mentioned binarized image series while there was one and only overlap of an organelle of interest between two consecutive images (see while-subloop). In this subloop, after thresholding the organelle of interest and finding its centroid in the $(l-1)^{\mbox{th}}$ labeled image, the position of this unique overlap including its number label was found in the $l^{\mbox{th}}$ labeled image. If there was found zero ($k = \emptyset$) or more than one unique overlap, the while looped was finished. The cause of disturbing the while-loop was two-fold:
\begin{itemize}
\item If $k = \emptyset$, an organelle did not occur in (vanished from) the $(l+1)^{\mbox{th}}$ image.
\item If number of elements in $k$ (i.e., if cardinality of the set $k$) is higher than 1, the organelle decayed in the $(l+1)^{\mbox{th}}$ image.
\end{itemize}

During the while-loop, the information about the time of saving was read from the names of each pair of original raw files (not illustrated in Fig.~\ref{Fig2}) which, after the subtraction, gave the scanning time $t_{scan,l}$ of the $l^{\mbox{th}}$ image.

After finishing the while-loop, the centroids saved from the relevant time-lapse images were plotted to obtain a trajectory of the moving organelle (Fig.~\ref{Fig4}--\textit{left}). The horizontal and vertical components of speed vectors were further computed as ratios of $x$- and $y$-components of length of trajectory's step to scanning times $t_{scan,l}$, respectively. Using the Pythagorean Theorem, we obtained the length of trajectory's step $d_l$ and of speed step. From these speed vectors, two resulted speed variables were calculated:
\begin{itemize}
\item Average speed characterizes an average path which an organelle travelled per a time unit:
\begin{equation}
AS = \frac{\sum_{l} (d_l \times px)}{\sum_{l}t_{scan,l}},
\label{eq2}
\end{equation} 
where
\begin{equation}
d_l = \sqrt{(\Delta x)_l^2 + (\Delta y)_l^2} 
\end{equation}
\item Speed resultant describes a total time-normalized path which an organelle travelled:
\begin{equation}
SR = \sum_l \left[\frac{px}{t_{scan,l}} \times (\Delta y)_l \right]
\label{eq4}
\end{equation}
\end{itemize}
In Eqs.~(\ref{eq2})--(\ref{eq4}), $px$ is a size of an image pixel ($px = 64$ nm). The differences $(\Delta x)_l$ and $(\Delta y)_l$ correspond to the subtraction of a position of the starting and ending point (i.e., pixel) in the $l^{\mbox{th}}$ and $(l+1)^{\mbox{th}}$ image at the $1^{\mbox{st}}$ and $2^{\mbox{nd}}$ coordinate, respectively.

Finally, these round values of speed vectors underwent the statistical evaluation -- namely, plotting the histogram of speed steps (in Sect.~\ref{supplementary}) and evaluation of the histogram's parameters (Tab.~\ref{Tab1}), plotting the trends of organelle's movements (Fig.~\ref{Fig4}--\textit{right}).
\end{enumerate}

\subsection{Verification of the Algorithm}\label{verification}

The above-described algorithm for analysis of movements of light-diffracting organelles was being verified on an unmodified MG-63 cell line which was scanned in bright-field mode using an inverted optical microscope (Sect.~\ref{methods}). The verification of the algorithm was performed by visual inspection through the time-lapse series (see Sect.~\ref{supplementary}). The segmented objects (organelles) were labeled by positions of centroids in Img. 1. The results of complete analysis are saved in Supplementary (Sect.~\ref{supplementary}).

\begin{figure}
\centering 
\includegraphics[width=.8\textwidth]{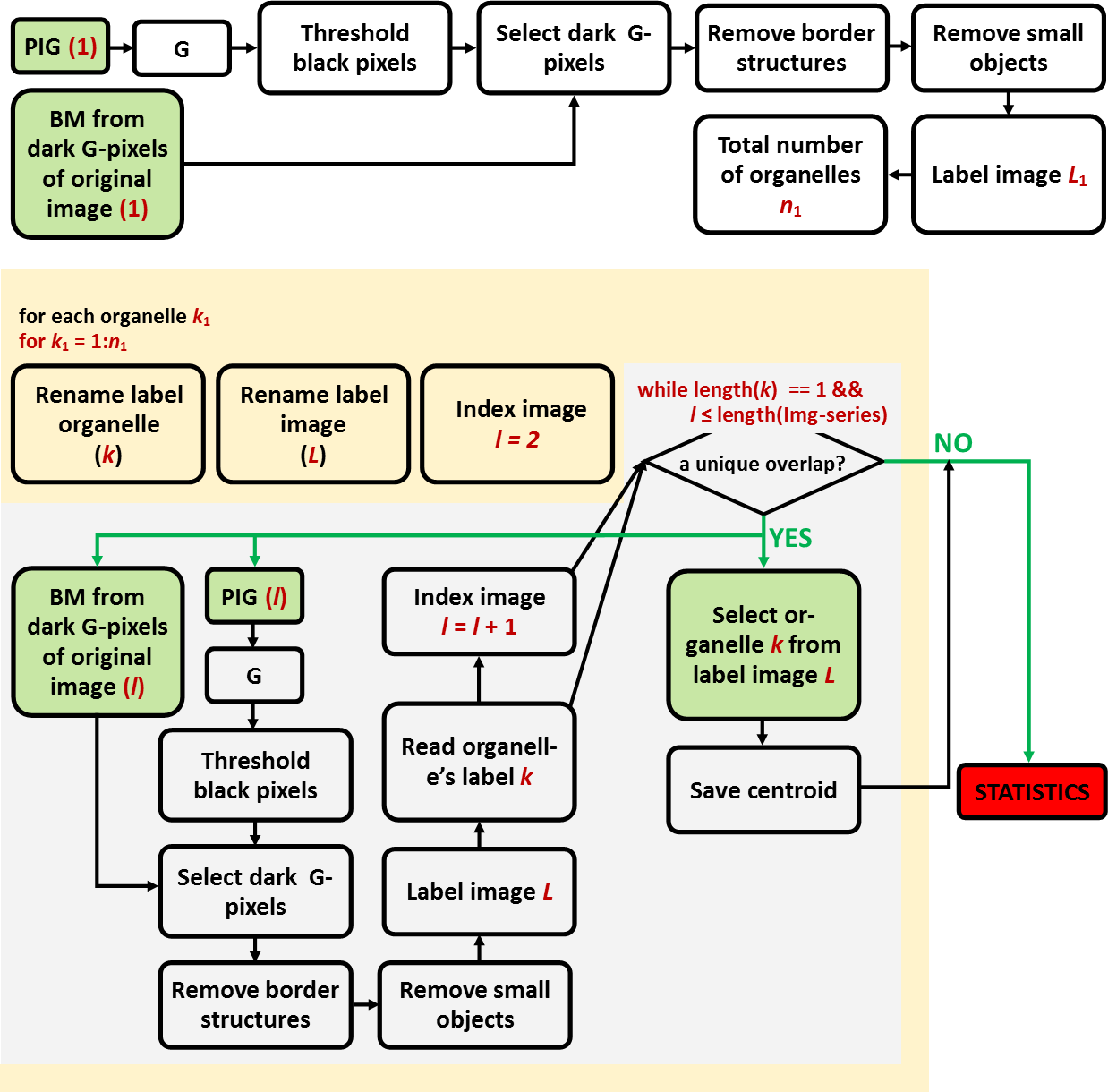}
\caption{Algorithm for organelles' tracking and statistical evaluation of movements inside an unlabelled live cell from time-lapse series obtained using wide-field bright-field photon transmission microscopy. \textit{white background} -- Segmentation of objects of interests from the 1$^{\mbox{st}}$ PIG-image, \textit{orange background} -- for-loop, \textit{gray background} -- while-loop, \textit{green boxes} -- inputs, \textit{red box} -- output.} 
\label{Fig2}
\end{figure}

In Img. 1, the algorithm detected 56 relevant objects, from which 46 objects were further found in Img. 2 and analyzed for the cause of disturbing the tracking:

\begin{enumerate}
\item Tracking of only one intracellular object (Organelle 4 at the position 095-118) was stopped since there was no overlap with the following image. In other words, the speed of the organelle was faster than the requested scanning speed of the camera.  
\item Finishing the tracking algorithm due to vanishing of organelle in the following image concerns to 12 organelles where re-appearing of the object in the next-next image occurred in 6 cases. This re-appearing occurred mainly due to the spectral properties of the observed organelle between series images which changes the possibility to detect the organelle using $\Gamma_{\alpha}^{(i)}$ computation.
\item 8 objects were split into more parts and the object in the next-to-last image was overlapped with more objects in the last image. This made the next tracking of the objects difficult. This splitting can be observed if either an intracellular object were divided into two (or more) objects (e.g., in case of division of mitochondria) or two objects were situated each above other and they moved mutually. In case of Organelle 3 (position 093-331), it is difficult to decide if there was its splitting or vanishing in the last image.
\item Only two objects (nucleoli Nos. 27 and 36 at the position 197-361 and 245-350, respectively) were detected through the whole 766-image series.
\item The observation of the rest of the objects was stopped due to joining a neighbouring organelle in the next-to-last image followed by re-splitting. Let us note that the joining of the objects shifts the centroid detected in its binary image and can distort the results of the followed statistics which is based on evaluation of trajectory and velocity steps.  
\end{enumerate}

\begin{figure}
\centering 
\includegraphics[width=\textwidth]{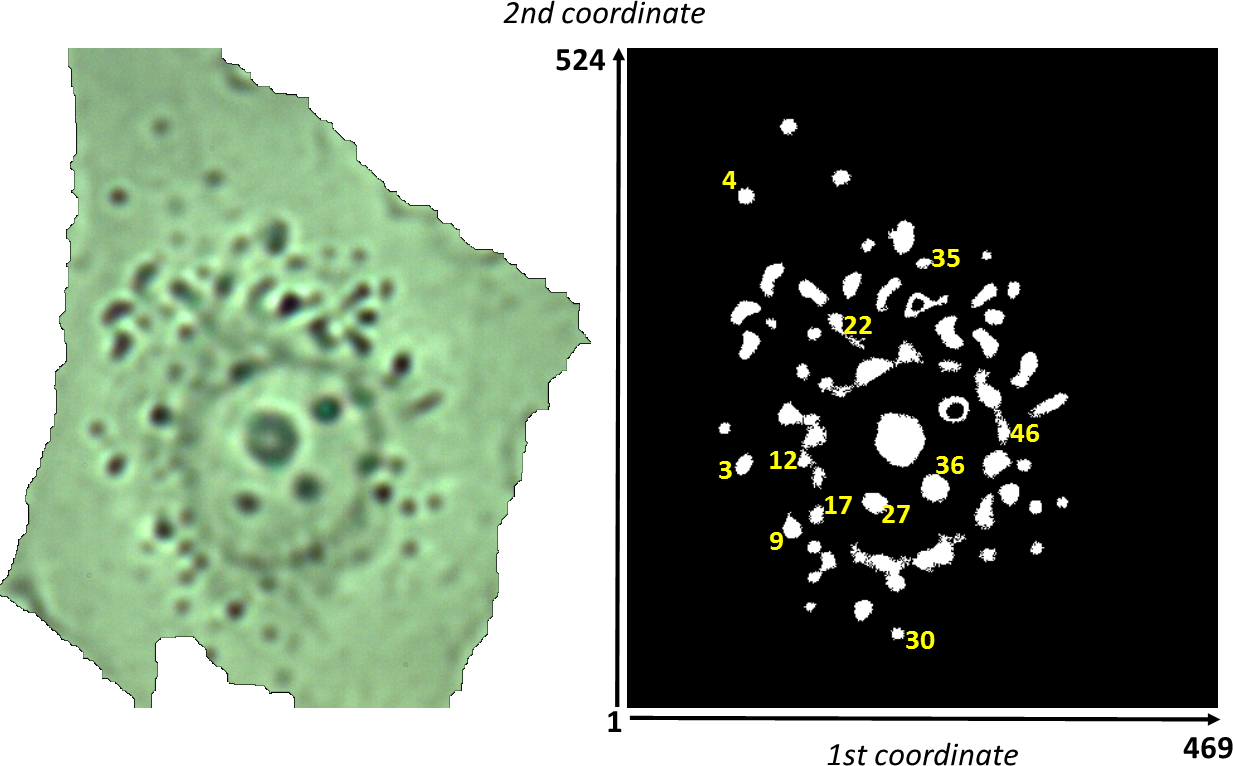}
\caption{Results of the segmentation algorithms. \textit{left} -- The 1$^{\mbox{st}}$ bright-field photon transmission time-lapse series image of an unmodified live MG-63 human osteosarcoma cell (visualized in 8-bit/c color depth). \textit{right} -- A binary image of organelles selected for statistical evaluation of their trajectories and speeds using the algorithm calculating the $\Gamma_{4.0}^{(i)}$. The coordinates describe the positions of the organelles as written in Supplementary. The Nos. of organelles correspond to the Nos. of organelles further in the text, figures, and Supplementary and are ordered according to the position of the organelle in the 1$^{\mbox{st}}$ binary image.}
\label{Fig3}
\end{figure}

The trajectories of these 46 trackable objects underwent further analysis (Sect.~\ref{algorithm}). Each shift of the trajectory step was normalized to time which gave the direction and the size of the vector of speed step. For each organelle, the statistical evaluation enables to assess the activity/stability of the movements as a standard deviation of the speed steps' sizes. These speed steps were further averaged by two ways: by average speeds (i.e., total speeds related to total times of tracking) and by speed resultants (which characterize trends of movements).

\begin{figure}
\centering 
\includegraphics[width=\textwidth]{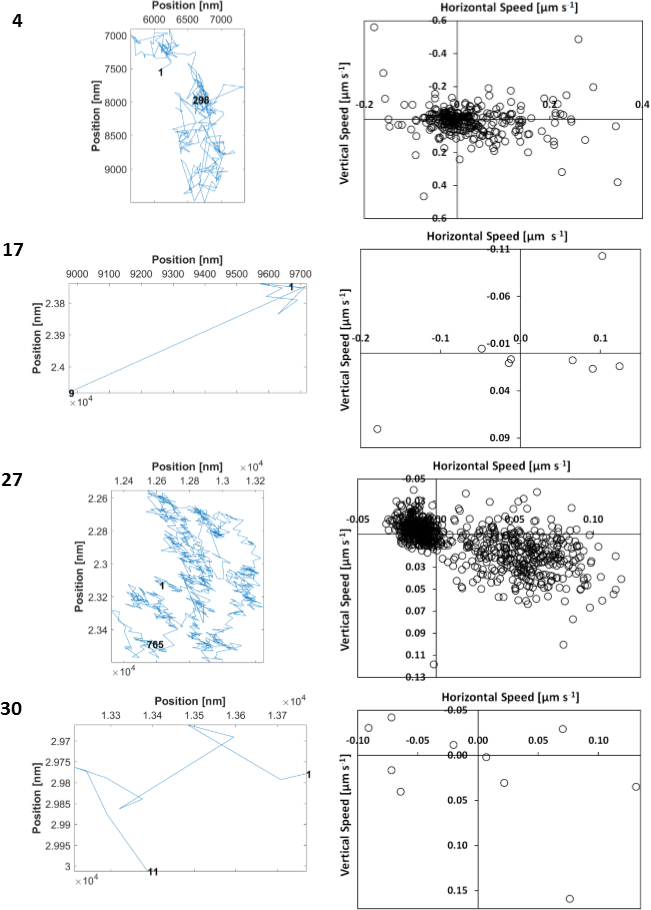}
\caption{The trajectories (\textit{left}) and the directions of speed steps (\textit{right}) of moving organelles inside of an unmodified live MG-63 human osteosarcoma cell. The Nos. of organelles correspond to the Nos. of organelles in the text, figures, and Supplementary and are ordered according to the position of the organelle in the 1$^{\mbox{st}}$ binary image. The Nos. of in the left graphs represent the first and the last time steps of tracking.} 
\label{Fig4}
\end{figure}

Examples of trajectories and directions of speed steps for four intracellular objects -- an active organelle moving in the cytosol (Organelle 4), an object embodied in the nuclear membrane (Organelle 17), a nucleolus (Organelle~27), and a small organelle in the cytosol (Organelle 30) -- are shown in Fig.~\ref{Fig4}--\textit{left}. Organelle~4 was tracked for 298 time steps with a high standard deviation of the speed steps of 86.023 nm s$^{-1}$ and a relatively high speed resultant of 8.374~$\mu$m~s$^{-1}$. In spite of the average value of average speed, this standard deviation and speed resultant make this organelle one of the most dynamic organelle in the intracellular space. Organelle 17 hit a neighbouring Organelle 9 (position 131-381) in time step 9 and separated in time step 10. The last detected time step 9 changed the statistics significantly by shifting the centroid of the binary image and, thus, by increasing the average speed. The standard deviation of the speed steps increase up to 63.976 nm s$^{-1}$ as well. However, the speed resultant (129.774 nm s$^{-1}$), and thus the total change of the position of Organelle 17, remained one of the lowest. In case of Organelle 27 (nucleolus), it is not surprised that it showed only a slight shaking movements during the whole time-lapse series. Its characteristics like the standard deviation of lengths of speed steps and average speed are of values of 29.605 nm s$^{-1}$, and 22.471~nm~s$^{-1}$, respectively. These results also correlate with the graph of directions of individual speed steps (Fig.~\ref{Fig4}--\textit{left}), where a substantial proportion of points is concentrated at the boundary of Quadrants II and III of the Cartesian plane with the deviations going to Quadrant I. However, the speed resultant is relatively high (17.652~$\mu$m~s$^{-1}$) which means that the nucleolus tended to move in one direction. The next studied intracellular object -- Organelle 30 -- vanished in time step 12, although it re-appeared in time step 13. This organelle showed average dynamics (the average speed of 52.513 nm s$^{-1}$) with a strongly below-average standard deviation of the length of speed steps (50.527 nm s$^{-1}$) and a low speed resultant (0.169 $\mu$m s$^{-1}$).

If we compare all organelles among each other and try to find the extremely behaving organelles, the least shaking organelle (standard deviation of lengths of time steps of 11.350 nm s$^{-1}$) with the lowest value speed resultant (7.778~$\mu$m~s$^{-1}$) is that of No. 35 (position 236-171), although it was trackable only through three time steps, despite it re-appeared in time step 5. The object with the highest standard deviation of length of time steps (1002.165 nm s$^{-1}$) is marked by No. 22 (position 171-222) and detected in four time steps. Nevertheless, in the 1$^{\mbox{st}}$ time step, this object seems to be formed by merging of two objects which were disconnected in the 2$^{\mbox{nd}}$ time step and one of the objects was joint to the other organelle. This led to the significant change of the position of the tracked centroid. Organelles 46 (a component of the nucleolar envelope) and 12 at the positions 290-282 and 141-327 has the lowest (7.115 nm s$^{-1}$) and the highest (284.381 nm s$^{-1}$) value of the average speed, respectively. However, in these cases, the resulted values have a low information value. Organelle 46 thus already split in time step 3, whereas Organelle 12 did not too move during the first two steps but it connected to a neighbouring object in time step 3.  

\begin{figure}
\centering 
\includegraphics[width=\textwidth]{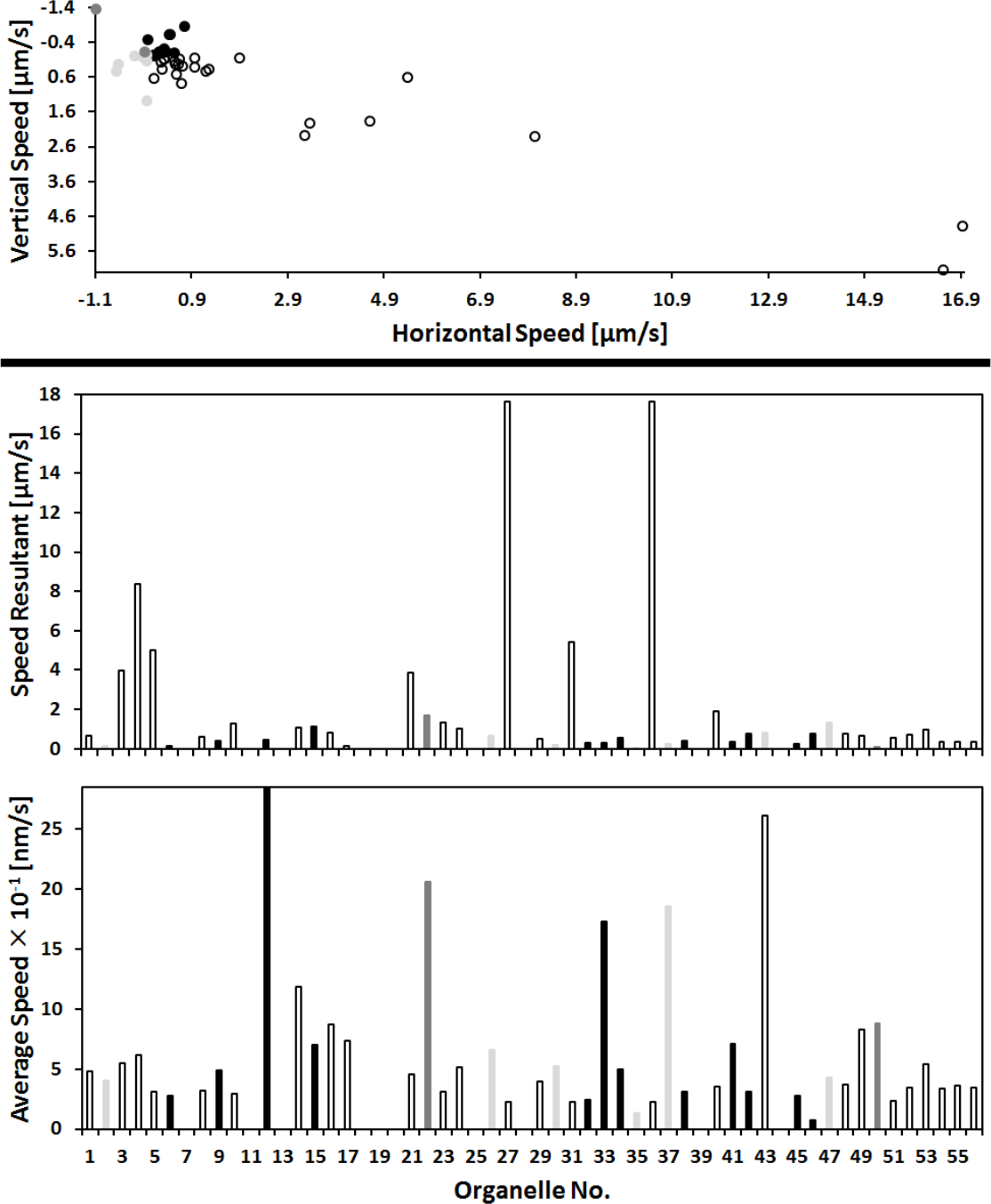}
\caption{The directions of speeds (\textit{upper}), the values of speed resultants (\textit{middle}), and the average speeds (\textit{lower}) of moving organelles inside of an unmodified live MG-63 human osteosarcoma cell. White, light gray, dark gray, and black bars represent the positions of the speed resultants in Quadrant I, II, III, and IV of the Cartesian plane, respectively. The Nos. of organelles correspond to the Nos. of organelles in the text, figures, and Supplementary and are ordered according to the position of the organelle in the 1$^{\mbox{st}}$ binary image. The positions without bars are attributed to non-trackable organelles which occurred only in Img. 1.} 
\label{Fig5}
\end{figure}

As seen in Fig.~\ref{Fig5}--\textit{upper}, only 7, 3, and 8 intracellular objects tended to move into Quadrant II, III, and IV of the Cartesian plane, respectively. The rest of objects were moving into Quadrant I (southeastern direction) where some attractor can be localized.

\begin{table}
\caption{Statistical evaluation of histograms of speed steps' vectors}
\label{Tab1}
\centering
\begin{tabular}{lllll}
\hline
    & \textbf{Mean} & \textbf{Median} & \textbf{Mode} & \textbf{STD} \\
    & \textbf{[nm s$^{-1}$]} & \textbf{[nm s$^{-1}$]} & \textbf{[nm s$^{-1}$]} & \textbf{[nm s$^{-1}$]} \\
\hline     
\textbf{Organelle 4} & 92.731 & 66.000 & 25.000 & 86.023 \\
\textbf{Organelle 17} & 87.875 & 78.500 & 14.000 & 63.976 \\ 
\textbf{Organelle 27} & 37.992 & 26.000 & 12.000 & 29.605 \\ 
\textbf{Organelle 30} & 78.500 & 76.000 & 76.000 & 50.527 \\
\hline
\end{tabular}
\end{table}

\section{Conclusions} \label{conclusions}

The presented automated label-free method based on image processing of bright-field photon microscopic time-lapse series provides an insight into the dynamical processes which can be crucial for understanding of many biological phenomena, including cell growth, mass transportation, signaling transduction and cell migration. The method showed a potential to be used, e.g., in pharmacology and toxicology for testing effects of chemical substances to tissue cultures. Nevertheless, the method is believed to be further improved by a possibility to identify the observed objects using calculation of a specific divergence derived from the R\'{e}nyi information entropy -- a point divergence gain -- which enables to specify spectral properties of an organelle by finding of pixels of unchanged intensities for two consecutive images in time-lapse series (if needed, fortified by a zig-zag z-stepping)~\cite{rychtarikova2016c}. The precision of the analysis can be increased by the usage of a camera of a higher scanning frequency and resolution.

\section{Methods} \label{methods}

\subsection{Cell cultivation}

A MG-63 cell line (Serva, cat. No. 86051601) was grown at low optical density overnight at 37$^\circ$C in a synthetic dropout media with 30\% raffinose as the sole carbon source. The nutrient solution for MG-63 cells consists of: 86\% EMEM, 10\% newborn-calf serum, 1\% antibiotics and antimycotics, 1\% L-glutamine, 1\% non-essential amino acids, 1\% $\mbox{NaHCO}_{3}$ (all components purchased from the PAA company). During microscopy experiment, cells were cultivated in a Bioptech FCS2 Closed Chamber System.

\subsection{Microscopy}\label{Microscopy}
Microscopy of a living MG-63 cell culture was performed using a versatile sub-microscope: a nanoscope developed for the Institute of Complex Systems FFPW by the company Optax Ltd. (Czech Republic). The optical path consisted of two Luminus 360 light emitting diodes, the condenser system, a firm sample holder, and an 40$\times$ objective system made of two complementary lenses that allow a change of distance between the objective lens and the sample. The UV and IR light emitted by LEDs was blocked by a 450-nm Longpass Filter and 775-nm Shortpass Filter (Edmund Optics), respectively. Next, a projective lens magnified the image to project on a Basler ACA2000-340kc camera chip of resolution of 2048$\times$1088 and 12-bit color depth. The size of the original camera pixel using primary magnification was 32$\times$32 nm$^2$. The combination of the time-lapse scan and zig-zag z-scan with a step of 300 nm was performed automatically with the average time step of (0.396 $\pm$ 0.594) fps to give a 766-image series of raw data around the focal plane obtained using a Bayer camera filter.

\section{Supplementary} \label{supplementary}

The Supplementary Data are stored at~\cite{ftp}.

\subsubsection*{Acknowledgments.} This work was supported by the Ministry of Education, Youth and Sports of the Czech Republic -- projects CENAKVA (No.\linebreak CZ.1.05/2.1.00/01.0024), CENAKVA II (No. LO1205 under the NPU I program), The CENAKVA Centre Development (No. CZ.1.05/2.1.00/19.0380) -- and by the CZ-A AKTION programme.

\end{document}